\documentclass[preprint,review,12pt]{elsarticle}
\usepackage{graphicx}
\usepackage{epsf}
\usepackage{amsmath}
\usepackage{lscape}
\usepackage{color}
\usepackage{dcolumn}
\usepackage{hyperref}
\usepackage{amsmath}
\usepackage{rotating}
\usepackage{soul}
\usepackage{longtable}
\usepackage{natbib}

\newcommand{\cm}{cm$^{-1}$}
\newcommand{\m}{$\mu$m}
\newcommand{\p}{$^\prime$}
\newcommand{\pp}{$^{\prime\prime}$}

\newcommand{\HtO}{H$_{2}$$^{16}$O}
\newcommand{\Hy}{H$_{2}$}

\newcommand{\etal}{\emph{et al}.}

\newcommand{\hto}{\hspace{2mm}}
\newcommand{\hsn}{\hspace{7mm}}
\newcommand{\hsx}{\hspace{6mm}}

\journal{Journal of Quantative Spectroscopy and Radiative Transfer}

\begin{document}
	
\begin{frontmatter}
		
\title{Pressure-dependent water absorption cross sections for exoplanets and other atmospheres.}
\author{Emma~J. Barton, C. Hill, Sergei~N. Yurchenko, Jonathan Tennyson}
\address{Department of Physics and Astronomy, University College London,
		London, WC1E 6BT, UK}
\author{Anna~S. Dudaryonok, Nina~N. Lavrentieva}
\address{V. E. Zuev Institute of Atmospheric Optics SB RAS, 1, Sq. Academician Zuev, 634021 Tomsk, Russia.}

\begin{abstract}

Many atmospheres (cool stars, brown dwarfs, giant planets,
extrasolar planets)
are predominately composed of molecular hydrogen and helium. \HtO\ is
one of the best measured molecules in extrasolar planetary atmospheres
to date and a major compound in the atmospheres of
brown-dwarfs and oxygen-rich cool stars, yet the
scope of experimental and theoretical studies on the pressure broadening
of water vapour lines by collision with hydrogen and helium remains limited.

Theoretical H$_2$- and He-broadening parameters of water vapour
lines (rotational quantum number $J$ up to 50) are obtained for
temperatures in the range 300 - 2000 K. Two approaches for calculation
of line widths were used: (i) the averaged energy difference method and
(ii) the empirical expression for $J$\p $J$\pp-dependence.

Voigt profiles based on these widths and the BT2 line list are used to
generate high resolution ($\Delta \tilde{\nu}$ = 0.01 \cm) pressure
broadened cross sections for a fixed range of temperatures and
pressures between 300 - 2000 K and 0.001 - 10 bar. An interpolation
procedure which can be used to determine cross sections at
intermediate temperature and pressure is described.
Pressure broadening parameters and cross sections are presented in
new ExoMol format.
	
\end{abstract}

\end{frontmatter}

\section{Introduction}

The ExoMol project aims to provide comprehensive line lists of
molecular transitions appropriate for molecules in hot atmospheres
such as those found in exoplanets, brown dwarfs and cool stars \cite{jt528}.
Until recently the ExoMol database did not include any information on the
pressure broadening of molecular lines despite the
known importance of pressure effects in these environments
\cite{jt521}. The scope of the ExoMol database is being
extended to systematically provide this additional data and other
supplementary data in order to maximise the usefulness of the line lists
\cite{jt631}. A major new feature is the inclusion, albeit at a fairly
crude level for most molecules, of pressure-broadening parameters
and pressure-dependent absorption cross sections.

\HtO\ is one of the few molecules detected in the atmosphere of an
exoplanet to date \cite{13TiEnCo.exo,15Crossfield.exo}. In particular
it has been identified in the atmospheres of several hot extrasolar
giant planets including HD189733b
\cite{jt400,10SwDeGr.exo,09SwVaTi.exo,13BiDeBr.exo,14McCrDe,14ToDeBu},
GJ 436b \cite{10StHaNy.exo,11KnMaCo.exo}, HD 209458b
\cite{jt488,13DeWiMc.exo}, XO-1b \cite{10TiDeSw.exo,13DeWiMc.exo},
HAT-P-1b \cite{13WaSiDe}, HD 179949b \cite{14BrDeBi.exo}, WASP-19b
\cite{13HuSiPo}, WASP-12b \cite{15KrLiBe} and planets orbitting HR
8799 \cite{13KoBaMa,15BaKoQu.exo}.  Test model calculations have shown
that inclusion pressure broadening of the water spectrum significantly
alters the observed spectrum of transiting hot Jupiter exoplanet
\cite{jt521}.

Prior to the discovery of exoplanets, \HtO\ was already known to be a
component of the atmospheres of cool stars and brown dwarfs
\cite{07ShBuxx.dwarfs,09Bexxxx.exo,12AlHoFr,15MoMaFo.dwarfs}, where it
can dominate the observed spectrum \cite{jt143}. Recent discoveries
have focused very cold brown dwarfs: Faherty \etal\
\cite{14FaTiSkMo.coolstar} made a tentative detection of water clouds
in WISE J085510.83-071442.5 and Skemer \etal\ \cite{16SkMoAl}

\HtO\ is one of the few molecules contained in the current release
of HITEMP \cite{jt480} which contains a comprehensive set of air-
and self-broadening parameters. These were obtained using a diet based
on physical principles and statistics \cite{07GoRoGa.broad} which
had a strong bias toward near-room temperatures.
As the bulk of the atmosphere of giant exoplanets and cool stars is
composed of a \Hy\ rich \Hy-He mix, molecular opacities in
atmospheric
models for these objects
should incorporate pressure broadening due to \Hy\ and He.

A number of works, reviewed in the next section, have provided
pressure broadening parameters for broadening of \HtO\ by \Hy\ and/or
He for this purpose, although the spectral, temperature and pressure
coverage is far from comprehensive.  Altogether, detailed information
for broadening of water vapour lines by \Hy\ and He is available for
around 1100 and 5000 lines respectively, while the most complete line
list for water to date (BT2, \cite{jt378}) contains around half a
billion lines which is sufficientlly complete for temperatures up to
3000 K. The temperature coverage, although appreciable for broadening
by \Hy\ (40 - 1500 K), is more limited for broadened by He (83 - 600
K).  Conversely the pressure coverage for broadening by He (up to 3
atm) is more extensive than for broadening by \Hy\ (up to 1330 mbar
$\approx$ 1.3 atm).

The first aim of this work is to provide \HtO-\Hy\ and \HtO-He
pressure broadening parameters suitable for temperatures up to 2000 K
and pressures up to 10 bar. These parameters can be used to generate a
pressure and temperature dependent Lorentzian half-width for every
line in the BT2 line list, or indeed any water line list with at least
rotational angular momentum ($J$\p-$J$\pp) quantum assignments.

The implementation of full line lists in atmospheric modelling codes,
although ideal, is often not practical due to their sheer size
\cite{jt542}. Hence the second aim of this work is to provide
pressure-dependant absorption cross sections based on
Voigt profiles for \HtO\ in a mixed
\Hy/He (85/15\%) environment for a range of temperatures (T = 300 -
2000 K) and pressures (P = 0.001 - 10 bar) relevant to exoplanet and
cool star atmospheres.

\section{Previous work}

For the \HtO-\Hy\ system the most extensive experimental study was performed
by Brown \& Plymate \cite{Brown1996263} who derived pressure broadened widths
for 630 lines in the range 55-4045 \cm\ at room temperature.
Other exclusively room temperature measurements have been carried out
by Steyert \etal\ \cite{Steyert2004183} (39 lines in the range 380-600\cm),
Brown \etal\ \cite{05BrBeDe.h2opb} (4 lines around 1540 \cm),
Golubiatnikov \cite{05Goxxxx.h2opb} (1 line at 183 GHz),
Lucchesinia, Gozzini \& Gabbanini \cite{00LuGoGa.h2opb} (15 lines in the range 820 - 830 nm),
and Zeninari \etal\ \cite{04ZePaCo.h2opb} (6 lines around 1.39\m).
The temperature and/or pressure dependence of the collision induced widths
for select (2 - 12) \HtO\ lines has been investigated by a handful studies
\cite{04ZePaCo.h2opb, Dick2009619, Langlois1994272, 93DuJoGo.h2opb}.
The works considering the widest range of temperatures and pressures are
those by Langlois, Birbeck \& Hanson \cite{Langlois1994272} and
Zeninari \etal\ \cite{04ZePaCo.h2opb} respectively. Langlois, Birbeck \& Hanson
combined two experimental set-ups, one incorporating a temperature controlled
static cell (limited to 450 K) the other a pressure driven shock tube, to
determine the temperature dependence of the pressure broadened widths of 12 \HtO\
lines over the range 300 - 1200 K with an expected accuracy of $\pm$25\%. The
dependence could be roughly represented by a power law of decreasing collision induced line width
with increasing temperature. Zeninari \etal\ made measurements of six \HtO\
lines at several pressures between 6 and 1330 mbar (ambient pressure $\approx$ 1000 mbar)
and concluded that the dependence could be described as a linear increase in collision
induced line width with pressure.

Theoretically determined pressure broadened widths and their temperature dependence
are available from Gamache, Lynch \& Brown \cite{Gamache1996471},
Faure \etal\ \cite{jt544} and Drouin \& Wisenfeld \cite{12DrWixx.h2opb}.
Drouin \& Wisenfeld focused on three lines at cold temperatures (below 200 K)
while Gamache, Lynch \& Brown and Faure \etal\ aimed to provide information for
hundreds of lines suitable for high temperature applications. The maximum
temperature considered by Faure \etal\ was hotter (1500 K vs 750 K) though
Gamache, Lynch \& Brown produced parameters for more lines (386 vs 228). All three
works focused on rotational transitions.

Several of the studies already mentioned
\cite{00LuGoGa.h2opb,Steyert2004183,05BrBeDe.h2opb,05Goxxxx.h2opb,Gamache1996471,Dick2009619,93DuJoGo.h2opb,04ZePaCo.h2opb}
also reported parameters for the \HtO-He system. Although the most
extensive studies of He-broadening coefficients of \HtO\ have been
performed by Petrova and co-workers
\cite{12PeSoSoSt.h2opb,13PeSoSoSt.h2opb,16PeSoSoSt.h2opb}, and Solodov
\& Starikov \cite{08SoStxx.h2opb,09SoStxx.h2opb}. Each of these works
presented measurements of pressure broadened widths for tens of lines
belonging to strong vibrational bands including $\nu_2 + \nu_3$
\cite{08SoStxx.h2opb,09SoStxx.h2opb,13PeSoSoSt.h2opb,16PeSoSoSt.h2opb},
$\nu_1 + \nu_2$ \cite{09SoStxx.h2opb,13PeSoSoSt.h2opb,16PeSoSoSt.h2opb},
$\nu_1 + \nu_2 + \nu_3$ \cite{12PeSoSoSt.h2opb,13PeSoSoSt.h2opb} and $2\nu_3$
and $\nu_1 + 2\nu_1$ \cite{16PeSoSoSt.h2opb} at room temperature and
varying pressures. All found the pressure dependence to be linear (up
to 3 atm \cite{13PeSoSoSt.h2opb,16PeSoSoSt.h2opb}). Petrova \etal\
\cite{13PeSoSoSt.h2opb} combined their own measurements with
literature data to determine the vibrationally and rotationally
dependent intermolecular potential for the \HtO-He system. This
potential was used by Petrova \etal\
\cite{13PeSoSoSt.h2opb,16PeSoSoSt.h2opb} to compute, and fit an
analytical expression to, helium pressure broadened width for
transitions belonging to 11 and 13 vibrational bands respectively with
rotational quantum numbers $J$ and $K_a$ up to 14 in the temperature
interval 83 $\leq T \leq $600 K. The temperature dependence in the
model was refined by comparison to previous temperature dependent
studies Goyette \& De Lucia \cite{90GoDexx.h2opb} (1 line at 183 GHz,
80 - 600 K) and Godon \& Bauer \cite{88GoBaxx.h2opb} (2 lines at 183
and 380 GHz, 300 - 390 K). In addition room temperature pressure
broadened widths for \HtO-He have been measured by Lazarev \etal\
\cite{95LaPnSu.h2opb} (1 line at 14397.4 \cm), Poddar \etal\
\cite{10PoMiHo.h2opb} (14 lines in the range 11,988-12,218 \cm),
Claveau \etal\ \cite{01ClHeHu.h2opb} (14 lines in the range 1850 -
2140 \cm) and Claveau \& Valentin \cite{09ClVaxx.h2opb} (10 lines in
the range 1170 - 1440 \cm).  Claveau \etal\ and Claveau \& Valentin
also investigated narrowing due to dynamic confinement (Dicke
narrowing \cite{Dixxxx.pb,WiDixx.pb}).

\section{Calculation of \Hy\ and He pressure induced line widths of \HtO\ spectral lines}

\subsection{Theoretical techniques used}

Two calculation techniques are used to determine the \HtO-\Hy\ and
\HtO-He Lorentzian half-widths. These techniques distinguish between
rigorous quantum numbers, namely total angular momentum $J$ and total
symmetry ($\Gamma_{\rm tot}$), which corresponds to parity and
ortho/para designation, and approximate projections of
the rotational motion ($K_a,K_c$). The techniques are:
\begin{enumerate}
	\item the averaged energy difference (AED) method \cite{14LaDuMa.h2opb} is used in cases where
	the complete set of quantum numbers (both rigorous and approximate) of a transition are known;
	\item the $J$\p $J$\pp- dependence technique \cite{10VoLaMi.h2opb} has been applied in cases where only
	rigorous (total angular momentum $J$ and total symmetry) were known.
\end{enumerate}
These approaches are described in detail in
the respective papers, thus only the main features are outlined below.

The averaged energy difference method allows the calculation of line widths of
asymmetric top molecules with approximately the same precision as in modern
theoretical and experimental methods without the need for a complicated calculation
scheme. The approach is based on matching the so called coupled energy state difference
with a line broadening value. Based on the modified Robert-Bonamy formalism, the expression
for the half-widths $\gamma_{if}$ is given by \cite{07MaTiBo.pb,06AnGaSz.pb}:

\begin{equation}
\gamma_{if} = \frac{n_b}{2\pi c} \int_{0}^{+\infty} v \, f(v)\, d v \int_{0}^{+\infty} 2 \, \pi\, b\, d b \left[1 - \cos(S_1 + \mathrm{Im} (S_2)) e^{-\mathrm{Re}(S_{2})}\right]
\end{equation}

\noindent
where indices $i$ and $f$ specify initial and final energy states,
$n_b$ is the number density of perturbers, $b$ is the impact parameter
and $v$ is the relative initial velocity and $f(v)$ is the Maxwell–Boltzmann distribution function
\cite{14LaDuMa.h2opb,07MaTiBo.pb,06AnGaSz.pb}. Real $S_1$ and complex
$S_2 = \mathrm{Re}(S_2) + i \mathrm{Im}(S_2)$ are the first and second order terms in
the expression for the scattering matrix.  These terms depend on
collision dynamics, the intermolecular potential and on the
ro-vibrational states of the molecule (and collision induced
transitions between these states). Expressions for $S_1$ and $S_2$ can
in found in Refs.
\cite{14LaDuMa.h2opb,98LyGaNe.h2opb,98GaLyNe.pb,95Lyxxxx.h2opb}.

The aim is to estimate how values of Re(S$_2$) vary with different
lines of interest.  Coupled energy state differences $\tilde{\omega}_{if}$ between an \HtO\
line of interest and other coupled lines ($\tilde{\omega}_{if}$ is defined explicitly
in \cite{14LaDuMa.h2opb}) are derived for thousands of lines. The
averaged energy differences for lines with experimentally-determined
collision-induced widths are then used to match averaged state energy
differences to line broadening values using a fitting formula.

\begin{figure}
	\begin{center}
		\scalebox{0.4}{\includegraphics{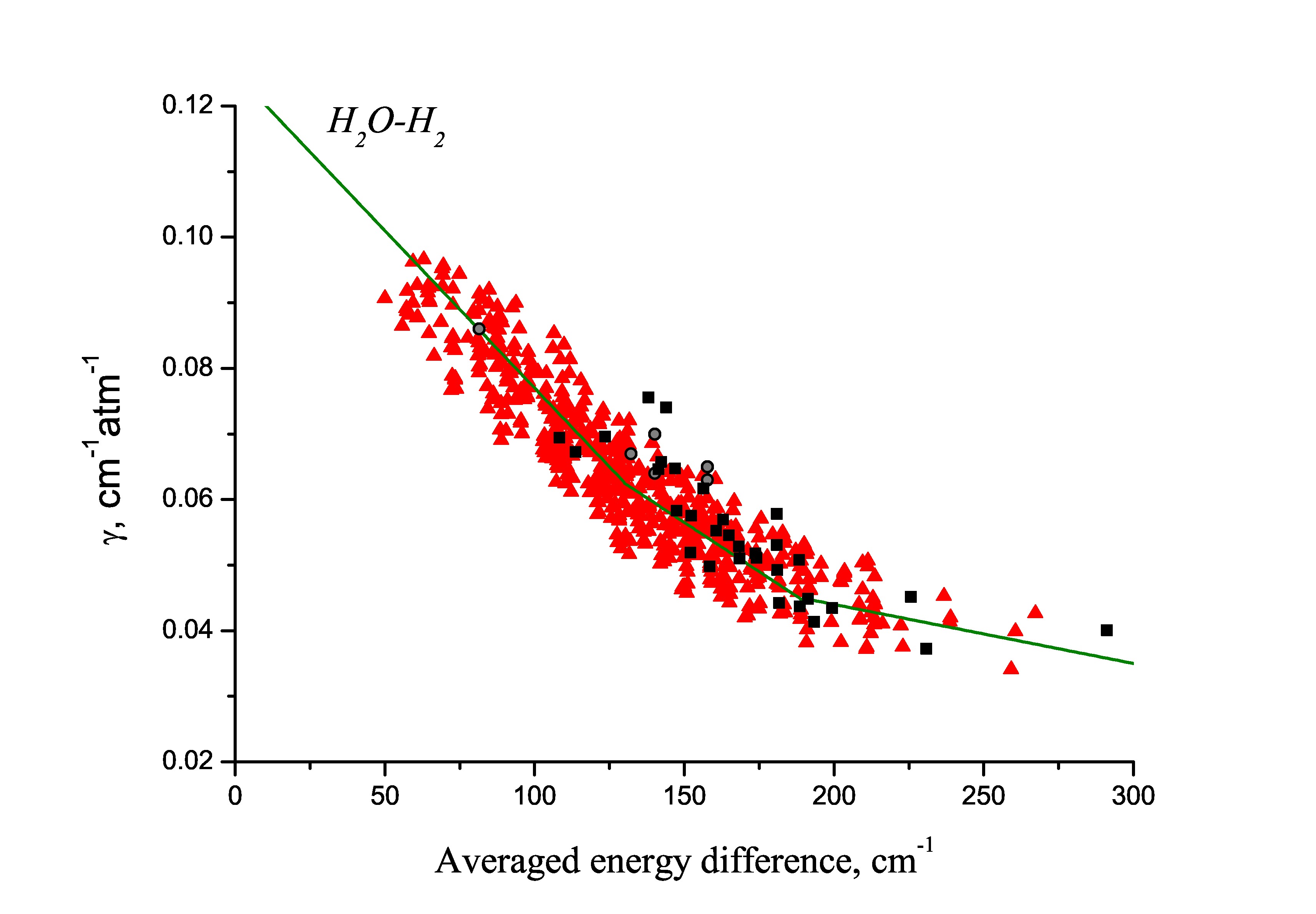}}
		\caption{Dependence of \HtO-\Hy\ broadening coefficients on averaged coupled energy state difference at 296 K; experimental data:
rotational band and fundamentals \cite{Brown1996263}, red triangles;
rotational band \cite{Steyert2004183}, black squares; $v_1+v_3$ and $2v_2$ \cite{Langlois1994272},
gray cirles.}
		\label{fig:avrE_model}
	\end{center}
\end{figure}

\begin{figure}
	\begin{center}
		\scalebox{0.4}{\includegraphics{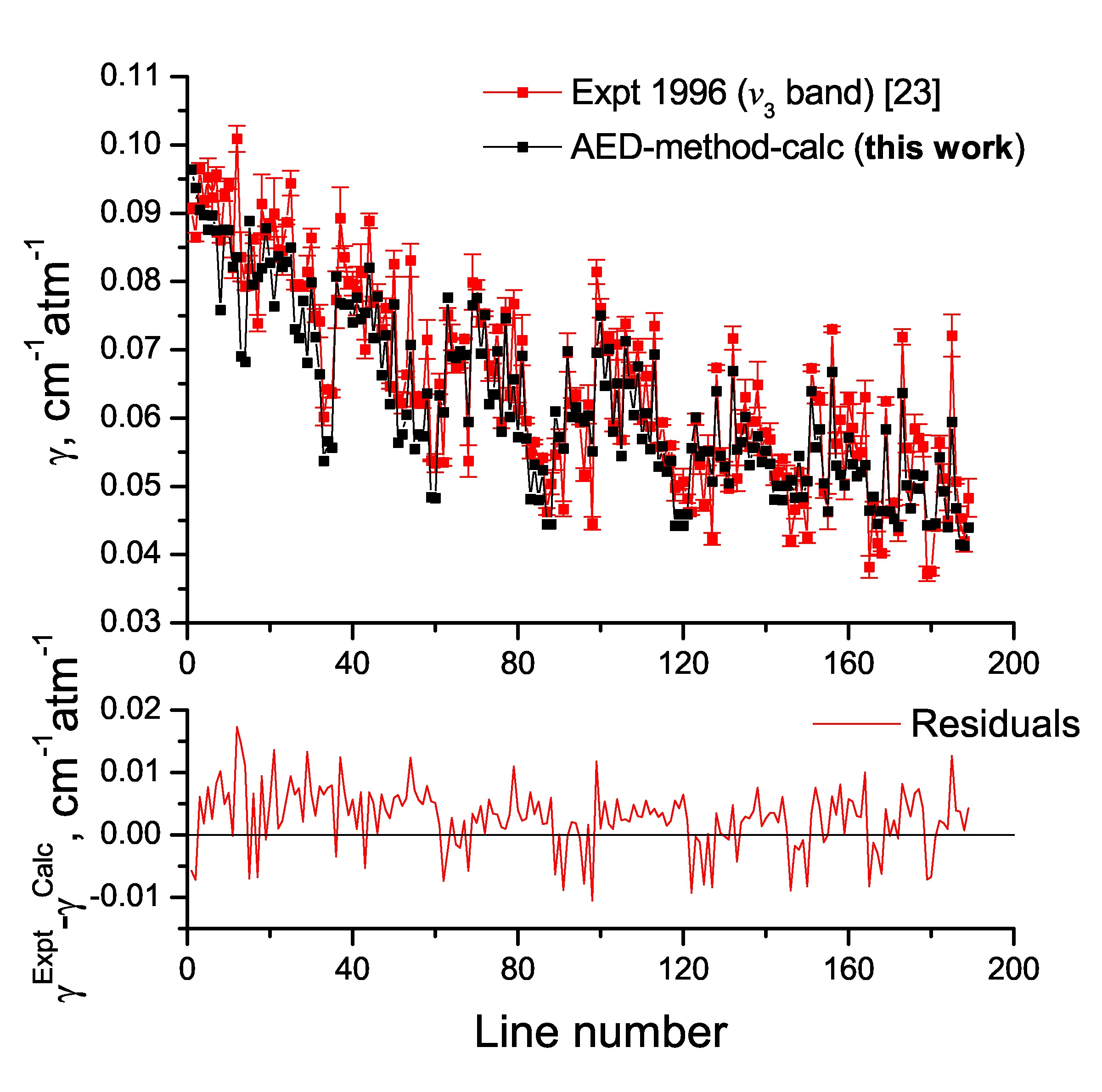}}
		\caption{Comparison of reconstructed \HtO-\Hy\ pressure dependent line widths with experimental values for the $v_3$ band \cite{Brown1996263} at 296 K.}
		\label{fig:avrE_comp_exp}
	\end{center}
\end{figure}

The dependence of broadening parameters on the averaged energy difference of
the coupled states is smooth. Figure~\ref{fig:avrE_model} illustrates this
for the \HtO-\Hy\ line widths. This figure demonstrates that one
can estimate corresponding widths with a simple fitting formula.
For other systems such N$_2$ broadening of water the observed data has
been found to lie almost exactly on a smooth curve \cite{jt432}; here
the scatter reflects, at least in part, uncertainty in the experimental data.
 The
reconstructed widths obtained from the fitting formula agree well
with the experimental values as shown in Figure~\ref{fig:avrE_comp_exp}
for \HtO-\Hy. This method requires assignment to normal modes (via quantum
numbers $v_1, v_2, v_3, K_a$ and $K_c$). However, these
quantum numbers are not always known or, indeed, always valid \cite{jt234}.

\begin{figure}
	\begin{center}
		\scalebox{0.4}{\includegraphics{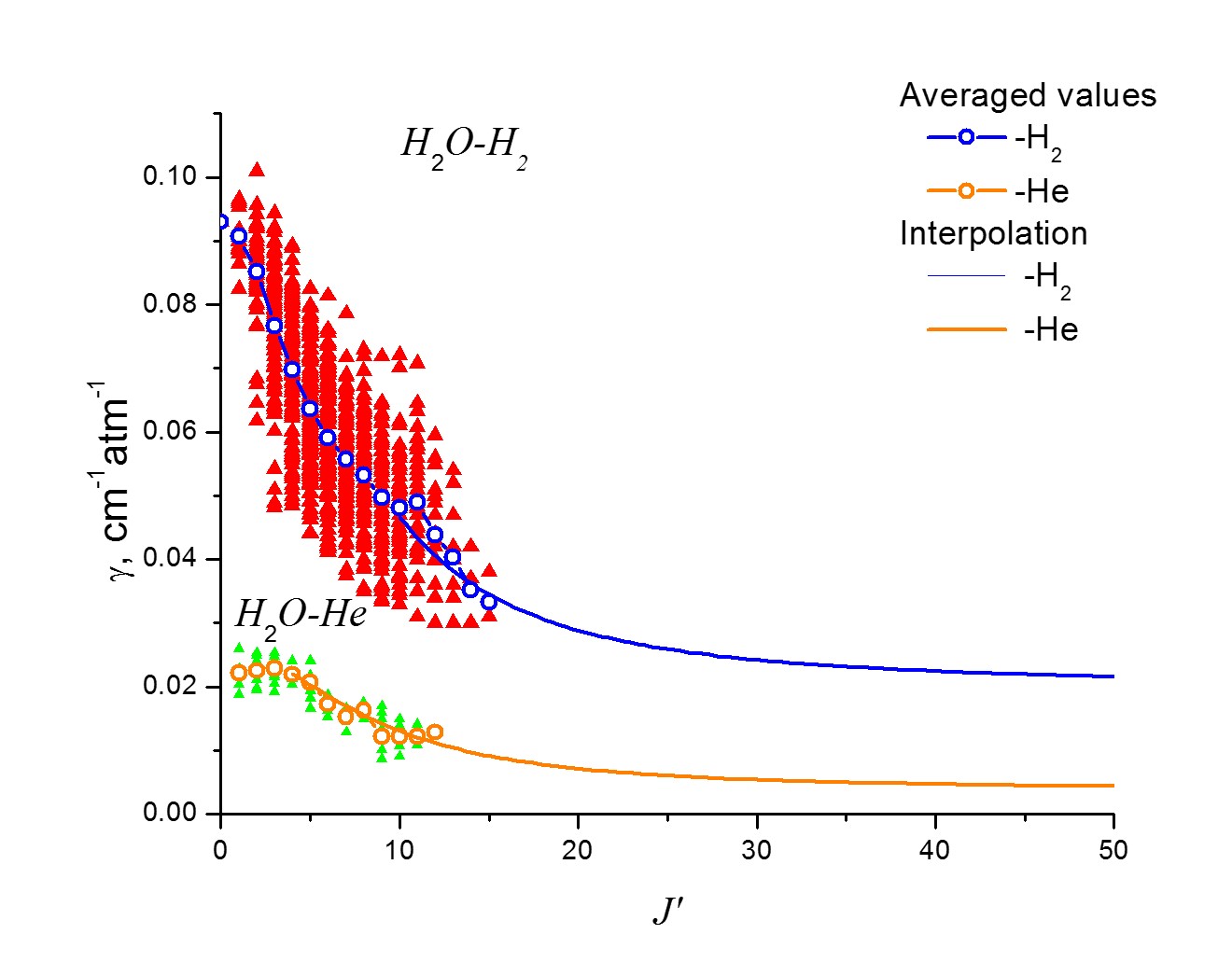}}
		\caption{Dependence of \HtO-\Hy\ and \HtO-He broadening coefficients on $J$\p\ at 296 K.}
		\label{fig:JJdep_model}
	\end{center}
\end{figure}

For cases where only good quantum numbers ($J$ and
$\Gamma_{\mathrm{tot}}$) are defined, the $J$\p - $J$\pp\ dependence
technique can be applied. Available widths are compiled and averaged
for each value of $J$\p\ per branch (P, Q and R) and total symmetry
(ortho/para). The resulting values are used to derive functions which
describe the dependence of the averaged width on $J$\p\ per branch and
per symmetry (if there is appreciable difference for the latter two).
This method applied to \HtO-\Hy\ and \HtO-He is illustrated in
Figure~\ref{fig:JJdep_model}.

\subsection{The \HtO-\Hy\ and \HtO-He line widths}

Both approaches described in the previous section were used to obtain
separate
\HtO-\Hy\ and \HtO-He Lorentzian half-widths for BT2 lines in the
range 500 - 10,000 \cm\ with an intensity that exceeds $1 \times 10^{-30}$ cm/molecule
in the temperature range 300 - 2000 K. Vibrational dependence is neglected
as it is found to change the value of the half-widths by only a few percent.
Dependence on $K_c$ and total symmetry was found to be much smaller than
dependence on $K_a$ and $J$\p-$J$\pp\ respectively.

Our calculated values range between 0.0208 - 0.0927 \cm\ atm$^{-1}$
for broadening by \Hy\ and 0.0043 - 0.0229 \cm\ atm$^{-1}$ for
broadening by He at 296 K. The average ratio between the \Hy\ to He
pressure induced line widths is 4.2. Broadening of water by H$_2$ is
much stronger than by He because the H$_2$ molecule has a quadrupole
moment. The electrostatic potential is given by an expansion of the
charge distribution in terms of the electric moments of the molecules.
So, the main contribution into H$_2$O -- H$_2$ broadening value is
from the dipole-quadrupole interaction, besides this term there is also a
quadrupole-quadrupole interaction contribution.  Helium does not have
any electric moments, so only the polarization potential (interactions
between induced moments) gives the contribution, which is very much weaker.

Tables~\ref{tab:H2comp} and \ref{tab:Hecomp} present a summary of the
comparison of our calculated values with available experimental and
previous theoretical studies.  The root mean square deviations (RMSD)
between this work and other works is given as a percentage of the
current values.  For the \HtO-\Hy\ system the RMSD is within 20\% or
the uncertainty quoted in the given reference for all studies. It
should be noted that for the comparison to Zeninari \etal\, the RMSD
of 19.1~\% is mainly down to the 6$_{60}$ $\leftarrow$ 6$_{61}$
transition which is in fact an unresolved doublet
\cite{04ZePaCo.h2opb}. The RMSD for the remaining five transitions is
8.4 \%.

For the \HtO-He system the RMSD is within 26 \% for all studies. Again for the comparison to Zeninari \etal\,
the RMSD of 15.0 \% is mainly down to the 6$_{60}$ $\leftarrow$ 6$_{61}$ transition.
The RMSD for the remaining five transitions is 8.6 \%.

\begin{table}
	\caption{Root mean square deviations (RMSD) of calculated values from available data on \HtO-\Hy broadening; $\gamma$ is the half-width and $n$ the
temperature exponent. In many cases there are no experimental data
available for comparison.}
	\label{tab:H2comp} \footnotesize
	\begin{center}
		\begin{tabular}{lrrr}
			\hline
			Reference                                              & Number of Transitions & RMSD $\gamma$ & RMSD $n$ \\
			\hline
			\textit{Experiment} & & & \\
			Steyert \etal\ \cite{Steyert2004183}                   & 39                    & 10.4 \%       & -        \\
			Brown \& Plymate \etal\ \cite{Brown1996263}            & 630                   &  7.4 \%       & -        \\
			Brown \etal\ \cite{05BrBeDe.h2opb}                     & 4                     &  8.8 \%       & -        \\
			Golubiatnikov \cite{05Goxxxx.h2opb}                    & 1                     & 13.9 \%       & -        \\
			Zeninari \etal\ \cite{04ZePaCo.h2opb}                  & 6                     & 19.1 \%       & -        \\
			Langlois, Birbeck \& Hanson \cite{Langlois1994272}     & 11                    & 17 \%         & 36.9 \%  \\
			Dutta \etal\ \cite{93DuJoGo.h2opb}                     & 2                     & 7.3 \%        & -        \\
			\textit{Theory} & & & \\
			Gamache, Lynch and Brown \cite{Gamache1996471}         & 386                   & 6.2 \%        &          \\
	                                                               & 32                    &               & 4.9 \%   \\
			Faure \etal\ \cite{jt544}                              & 228                   & 24.1 \%       & 18.2 \%  \\
 			\hline
		\end{tabular}
	\end{center}	
\end{table}

\begin{table}
	\caption{Root mean square deviations (RMSD) of calculated values from available data on \HtO-He broadening; $\gamma$ is the half-width and $n$ the
temperature exponent.}
	\label{tab:Hecomp} \footnotesize
	\begin{center}
		\begin{tabular}{lrrr}
			\hline
			Reference & Number of Transitions & RMSD $\gamma$ & RMSD $n$ \\
			\hline
			\textit{Experiment} & & & \\
			Petrova \etal\ \cite{16PeSoSoSt.h2opb}                 & 103                   & 17.1 \%       &  -              \\
			Petrova \etal\ \cite{13PeSoSoSt.h2opb}                 & 150                   & 17.7 \%       &  -              \\
			Petrova \etal\ \cite{12PeSoSoSt.h2opb}                 & 105                   & 15.5 \%       &  -              \\
			Solodov \& Starikov \cite{08SoStxx.h2opb}              & 32                    & 14.0 \%       &  -              \\
			Solodov \& Starikov \cite{09SoStxx.h2opb}              & 53                    & 13.9 \%       &  -              \\
			Goyette \& De Lucia \cite{90GoDexx.h2opb}              & 1                     & 1.7 \%        &  42.2 \%        \\
			Godon \& Bauer \cite{88GoBaxx.h2opb}                   & 2                     & 10.3 \%       &  13.6 \%        \\
			Poddar \etal\ \cite{10PoMiHo.h2opb}                    & 14                    & 14.0 \%       &  -              \\
			Claveau \& Valentin \cite{09ClVaxx.h2opb}              & 10                    &  25.7 \%      &  -              \\
			Claveau \etal\ \cite{01ClHeHu.h2opb}                   & 14                    & 24.7 \%       &  -              \\
			Steyert \etal\ \cite{Steyert2004183}                   & 39                    & 23.3 \%       &  -              \\
			Brown \etal\ \cite{05BrBeDe.h2opb}                     & 4                     & 11.1 \%       &  -              \\
			Dutta \etal\ \cite{93DuJoGo.h2opb}                     & 2                     & 9.2 \%        &  -              \\
			Zeninari \etal\ \cite{04ZePaCo.h2opb}                  & 6                     & 15.0 \%       &  -              \\
			Golubiatnikov \cite{05Goxxxx.h2opb}                    & 1                     & 4.0 \%        &  -              \\
			Lazarev \etal\ \cite{95LaPnSu.h2opb}                   & 1                     & 7.1 \%        &  -              \\
			\textit{Theory} & & & \\
			Gamache, Lynch and Brown \cite{Gamache1996471}         & 386                   & 6.2 \%        &                 \\
			                                                       & 32                    &               & 4.9 \%          \\
			\hline
		\end{tabular}
	\end{center}	
\end{table}

\subsection{Temperature and pressure dependence of the \HtO-\Hy/He line widths}

In order to determine the temperature dependence, the half-width
calculations were made at the following temperatures:
$T$ = 300, 400, 500, 600, 700, 800, 900, 1000, 1200, 1400, 1600,
1800, 2000~K. The temperature dependence,
represented by exponent $n$, was obtained by fitting to the standard relation:
\begin{equation}
\label{eq:Tdep}
\gamma(T) = \gamma_{\textrm{ref}} \times \left(\frac{T_{\textrm{ref}}}{T}\right)^{n}.
\end{equation}
\noindent
where $T_{\textrm{ref}}$ = 296 K is the reference temperature,
$\gamma_{\textrm{ref}}$ is the Lorentzian half-width at reference
temperature and $n$ is the temperature exponent. The temperature
exponents vary from 0.866 to 0.027 for broadening by \Hy\ and from
0.5 to 0.02 for broadening by He. Comparisons with experimentally
derived and previously calculated values are summarised in Tables~\ref{tab:H2comp}
and \ref{tab:Hecomp}.

Recently Wilzewski \etal\ \cite{16WiGoKo.pb}
commented that a single power law of the form Eq. (\ref{eq:Tdep}) only works well
within relatively narrow temperature intervals. However they also note
that there is not enough experimental data to characterise an alternative
model at present, and we find Eq. (\ref{eq:Tdep}) reproduces our calculated
values for the temperature range 300 - 2000 K to sufficient accuracy
(within 3\%). For high accuracy treatments other issues arise with use of the
Voigt profile \cite{jt584}.

The pressure dependence of the Lorentzian half-widths in the range
0.001 - 10 bar is assumed to be linear:
\begin{equation}
\label{eq:P_dep}
\gamma(P) = \gamma_{\textrm{ref}} \times \left(\frac{P}{P_{\textrm{ref}}}\right).
\end{equation}
where $P_{\textrm{ref}}$ = 1 bar is the reference pressure. This is
indicated by available pressure dependent experimental investigations
up to around ambient pressure for broadening by \Hy\ \cite{04ZePaCo.h2opb},
and up to around 3 atmospheres ($\approx$ 3 bar) for broadening by He
\cite{13PeSoSoSt.h2opb,16PeSoSoSt.h2opb}. Although measurements at the
high end of the pressure range ($\sim$ 10 bar) would be helpful to verify
this; in particular high pressure leads to three-body effects which are
implicitly neglected in our formulation.

\subsection{The .broad files}

In principle parameters were calculated for around 4 million BT2 lines
with an intensity that exceeds $1 \times 10^{-30}$ cm/molecule
in the temperature range 300 - 2000 K. In practice, since vibrational
dependence was neglected and dependence on total symmetry and rotational
quantum number $K_c$ was found to be comparatively small, the results comprise
of around 23,000 widths that depend on $J$\p, $J$\pp, $K_a$\p and $K_a$\pp,
and 100 widths that depend on $J$\p\ and $J$\pp\ only. This can be
represented in new ExoMol format \cite{jt631} as two \verb|.broad| files:
\newline \verb|1H2-16O__H2__NLAD.broad|, extract shown in Table~\ref{tab:broad_extract_H2}
\newline \verb|1H2-16O__He__NLAD.broad|, extract shown in Table~\ref{tab:broad_extract_He}
\newline These files and Eq.(\ref{eq:TPE_dep}) below can be used to generate
a pressure and temperature dependent Lorentzian half-width for any water
vapour line, with at least $J$\p - $J$\pp\ quantum assignments, for a pure
\Hy, pure He or mixed \Hy/He environment.

\begin{equation}
\label{eq:TPE_dep}
\gamma(T,P) = \left(\frac{T_{\textrm{ref}}}{T}\right)^{n} \times \left(\frac{P}{P_{\textrm{ref}}}\right) \times \sum_{b} \gamma_{\textrm{ref,b}} p_b.
\end{equation}

\noindent
where $\gamma_{\textrm{ref,b}}$ is the Lorentzian half-width due to a
specific broadener in units \cm/bar (ExoMol convention) and $p_b$ is the partial pressure
of the broadener. Note that to convert from \cm/bar used by ExoMol to \cm/atm
used by Hitran requires $\gamma$ to be mutiplied by 1.01325.

The \verb|.broad| files presented as part of this work differ slightly
from those included in the ExoMol database. The \verb|.broad| files
provided with the ExoMol line lists, described in \cite{jt631},
contain data from multiple sources while the \verb|.broad| files
available as supplementary data to this work contain data from only
one source, the current work. Hence, the file names are appended with
a dataset name NLAD. We recomend the use of the full \verb|.broad|
file presented on the ExoMol website as this includes both
line-by-line parameters where available, which are essential for
detailed and high resolution studies, while the current model which
allows a width to be generated for every molecular line.

\begin{table}
	\caption{\texttt{1H2-16O\_\_H2\_\_NLAD.broad}: Extract of \HtO-\Hy\ broad file: portion of the file (upper part); field specification (lower part).}
	\label{tab:broad_extract_H2} \footnotesize
	\begin{center}
		\begin{tabular}{lllllll}
			\hline
			a3 & 0.0690 & 0.502 & \hsn5 & \hsn4 & \hto1 & \hto3 \\
			a3 & 0.0580 & 0.424 & \hsn6 & \hsn6 & \hto3 & \hto3  \\
			a3 & 0.0754 & 0.541 & \hsn3 & \hsn3 & \hto0 & \hto1  \\
			... & & & & & &  \\			
			a1 & 0.0342 & 0.253 & \hsx14 & \hsx15 && \\
			a1 & 0.0328 & 0.241 & \hsx15 & \hsx16 && \\
			a1 & 0.0317 & 0.240 & \hsx16 & \hsx17 && \\
			... & & & & & & \\
		\end{tabular}
		\begin{tabular}{llll}
			\hline
			Field & Fortran Format & C format & Description \\
			\hline
			code & A2 & $\%$2s & Code identifying quantum number set following $J$\pp* \\
			$\gamma_{\textrm{ref}}$ & F6.4 & \%6.4f & Lorentzian half-width at reference temperature and pressure in \cm/bar \\
			$n$ & F5.3 & \%5.3f & Temperature exponent \\
			$J$\pp & I7 & $\%$7d & Lower $J$-quantum number \\
			$J$\p & I7 & $\%$7d & Upper $J$-quantum number \\
			$K_a$\pp & I2 & $\%$2d & Lower rotational quantum number \\
			$K_a$\p & I2 & $\%$2d & Upper rotational quantum number \\
			\hline
		\end{tabular}
	\end{center}
	\noindent
	*Code definitions:
	a3 = parameters presented as a function of $J$\pp (compulsory) and $J$\p, $K_a$\pp and $K_a$\p.
	a1 = parameters presented as a function of $J$\pp (compulsory) and $J$\p.
\end{table}

\begin{table}
	\caption{\texttt{1H2-16O\_\_He\_\_NLAD.broad}: Extract of \HtO-He broad file: portion of the file (upper part); field specification (lower part).}
	\label{tab:broad_extract_He} \footnotesize
	\begin{center}
		\begin{tabular}{lllllll}
			\hline		
			a3 & 0.0213 & 0.269 & \hsn5 & \hsn4 & \hto1 & \hto3 \\
			a3 & 0.0183 & 0.241 & \hsn6 & \hsn6 & \hto3 & \hto3  \\
			a3 & 0.0226 & 0.320 & \hsn3 & \hsn3 & \hto0 & \hto1  \\
			... & & & & & &  \\			
			a1 & 0.0092 & 0.179 & \hsx14 & \hsx15 && \\
			a1 & 0.0086 & 0.145 & \hsx15 & \hsx16 && \\
			a1 & 0.0082 & 0.148 & \hsx16 & \hsx17 && \\
			... & & & & & & \\
		\end{tabular}
		\begin{tabular}{llll}
			\hline
			Field & Fortran Format & C format & Description \\
			\hline
			code & A2 & $\%$2s & Code identifying quantum number set following $J$\pp* \\
			$\gamma_{\textrm{ref}}$ & F6.4 & \%6.4f & Lorentzian half-width at reference temperature and pressure in \cm/bar \\
			$n$ & F5.3 & \%5.3f & Temperature exponent \\
			$J$\pp & I7 & $\%$7d & Lower $J$-quantum number \\
			$J$\p & I7 & $\%$7d & Upper $J$-quantum number \\
			$K_a$\pp & I2 & $\%$2d & Lower rotational quantum number \\
			$K_a$\p & I2 & $\%$2d & Upper rotational quantum number \\
			\hline
		\end{tabular}
	\end{center}
	\noindent
	*Code definitions:
	a3 = parameters presented as a function of $J$\pp (compulsory) and $J$\p, $K_a$\pp and $K_a$\p.
	a1 = parameters presented as a function of $J$\pp (compulsory) and $J$\p.
\end{table}

\section{Computation of \HtO\ absorption cross-sections}

\subsection{Method}

The high resolution cross section is calculated on an evenly spaced wavenumber
grid, $\tilde{\nu}_i$, defining bins of width $\Delta \tilde{\nu}$. A Voigt profile is
used to model the joint contributions from thermal and collision induced
broadening:

\begin{equation}
f_{V}(\tilde{\nu}, \tilde{\nu}_{0;j}, \alpha_j, \gamma_j) = \frac{1}{\sqrt{\pi}} \frac{\sqrt{\ln 2}}{\alpha_j} \mathrm{wofz}\left(\frac{\tilde{\nu} - \tilde{\nu}_{0;j}}{\alpha_j}\sqrt{\ln 2} + i\frac{\gamma_j}{\alpha_j}\sqrt{\ln 2}\right)
\end{equation}

\noindent
where wofz is the scaled complex complementary error function, also known
as the Faddeeva function. This is calculated using the Faddeeva package 
\cite{faddeeva}.
$\tilde{\nu}_{0;j}$ is the line centre, $\gamma_j$ is the Lorentzian half-width
at half-maximum and $\alpha_j$
is the Doppler half-width at half-maximum given by:

\begin{equation}
\alpha = \sqrt{\frac{2 k T \ln 2}{m}} \frac{\tilde{\nu}_{0;j}}{c}
\end{equation}

\noindent
at temperature $T$ in K for a molecule of mass $m$ in kg. Note that in
the limit $\alpha \gg \gamma$, the profile reduces to a Gaussian:

\begin{equation}
f_{G}({\tilde{\nu}}, \tilde{\nu}_{0;j}, \alpha_j) = \sqrt{\frac{\ln 2}{\pi}} \frac{1}{\alpha_j}
\exp\left(-\frac{(\tilde{\nu} - \tilde{\nu}_{0;j})^{2}\ln 2}{\alpha_j^{2}}\right)
\end{equation}

\noindent
whereas for $\alpha \ll \gamma$ the profile reduces to a Lorentzian:

\begin{equation}
f_{L}({\tilde{\nu}}, \tilde{\nu}_{0;j}, \gamma_j) = \frac{\gamma_j}{\pi}\frac{1}{(\tilde{\nu} - \tilde{\nu}_{0;j})^2 + \gamma_j^2}
\end{equation}

\noindent
However a Voigt profile is evaluated for every absorption line in all calculations of
cross sections presented in this work. The cross section for each bin is the sum of
the contributions from individual lines:

\begin{equation}
\sigma_{i} = \sum_{j} \sigma_{ij}
\end{equation}

\noindent
where:

\begin{equation}
\label{eq:pointwise}
\sigma_{ij} = S_{j} f_{V}(\tilde{\nu}, \tilde{\nu}_{0;j}, \alpha_j, \gamma_j).
\end{equation}

\noindent
$S_{j}$ is the line intensity in units of cm per molecule given by:

\begin{equation}
S_{j} = \frac{A_{j}}{8 \pi c} \frac{g_{j}^{\prime} \exp\left({-c_{2}E_{j}^{\prime\prime}/T}\right)}{\tilde{\nu}_{0;j}^{2} Q(T)} \left(1 - \exp\left(-\frac{c_{2}\tilde{\nu}_{0;j}}{T}\right)\right)
\end{equation}

\noindent
Here, $g_{j}^{\prime}$ and $E_{j}^{\prime\prime}$ are the total upper-state degeneracy and
lower-state energy respectively, $A_{j}$ is the Einstein A coefficient for the
transition and $c_{2} = hc/k_{B}$ is the second radiation constant. $\tilde{\nu}_{0;j}$ and
$A_{j}$ were taken from the BT2 line list while the molecular partition function, Q(T), was obtained
from the tabulated values of Vidler \& Tennyson \cite{jt263}.

When evaluating a Voigt profile it is necessary to select an
appropriate profile grid resolution and evaluation width in order to
adequately sample the contribution from the profile whilst considering
the computational cost.  For the profile grid resolution we adopted
the staggered wavenumber grid spacings of \cite{jt542} (see
Table~\ref{tab:gridv}). These spacings are well below the Voigt width
for the pressure and temperature range considered, which is essential
for Eq. (\ref{eq:pointwise}) to be valid.

For the profile evaluation width we adopted a cut-off of 200 $\times (\alpha_j + \gamma_j)$ either
side of the centroid. We find that this is sufficient to capture the contributions from the wings
in a manner that adapts with both pressure and temperature.

We note Hedges \& Madhusudhan \cite{16HeMaxx} recently
investigated the effect of various factors, including the profile grid
resolution and evaluation width, on pressure dependent absorption
cross sections for application to exoplanet atmospheres. Hedges \&
Madhusudhan proposed a more computationally-efficient grid resolution,
a pressure adaptive grid, than the staggered grid employed by this
work. However, they  comment that the pressure adaptive grid
resulted in a small loss of opacity ($\sim$ 2\%) compared to the
staggered grid at the high end of the pressure range (P $\sim$ 10
bar). Hence we opted for the staggered grid.  For the profile
evaluation width Hedges \& Madhusudhan compared two methods, a
pressure adaptive fixed cut-off and a cut-off evaluated per line as a
multiple of Lorentzian half-widths, to their approach, a cut-off
evaluated per line as a multiple of effective Voigt half-widths.  Their
effective Voigt half-width was defined as \cite{16HeMaxx}:

\begin{equation}
\gamma_V \approx 0.5346\gamma_L \sqrt{0.2166\gamma_L^2 + \gamma_G^2}
\end{equation}

\noindent
where $\gamma_G = \alpha_j$, $\gamma_L = \gamma_j$ and $\gamma_V < (\alpha_j + \gamma_j)$ for
comparison to this work. Hedges \& Madhusudhan found that both the fixed cut-off method and
Lorentzian half-width analysis resulted in significant ($>$ 40\%) opacity loss at the low end
of the pressure range (P $<$ 0.01 bar) compared to the Voigt half-width analysis. As our approach
is very similar to Hedges \& Madhusudhan we should also avoid this loss of opacity.

\begin{table}
	\caption{Summary of the grid spacings for the cross sections calculated in difference wavenumber regions}
	\label{tab:gridv} \footnotesize
	\begin{center}
		\begin{tabular}{lc}
			\hline
			Wavenumber range \cm & Grid spacing \cm \\
			\hline
			10 - 100       &       10$^{-5}$        \\
			100 - 1000      &       10$^{-4}$             \\
			1000 - 10,000      &       10$^{-3}$              \\
			10,000 - 30,000         &        10$^{-2}$ \\
			\hline
		\end{tabular}
	\end{center}	
\end{table}

\subsection{Results}

Pressure dependent absorption cross sections for \HtO\ in a mixed \Hy/He
environment (85/15\%\ by number) are calculated on a fixed temperature
and pressure grid (see Table~\ref{tab:gridTP}) using the BT2
line list, pressure broadening parameters determined as part of this work (Section 1)
and the method described above. However, no attempt is made include contributions
to the opacity from the water vapour continuum or water dimer absorption.

The cross sections were calculated between 10 and 30,000 \cm\ using the staggered
wavenumber grid given in Table~\ref{tab:gridv} and then binned to a common grid
spacing of 0.01 \cm. Each region was calculated to overlap with its neighbours
by at least 1 \cm, which we find to be sufficient to avoid discontinuities when
they are binned to a common grid spacing.

\begin{figure}
	\begin{center}
		\scalebox{0.4}{\includegraphics{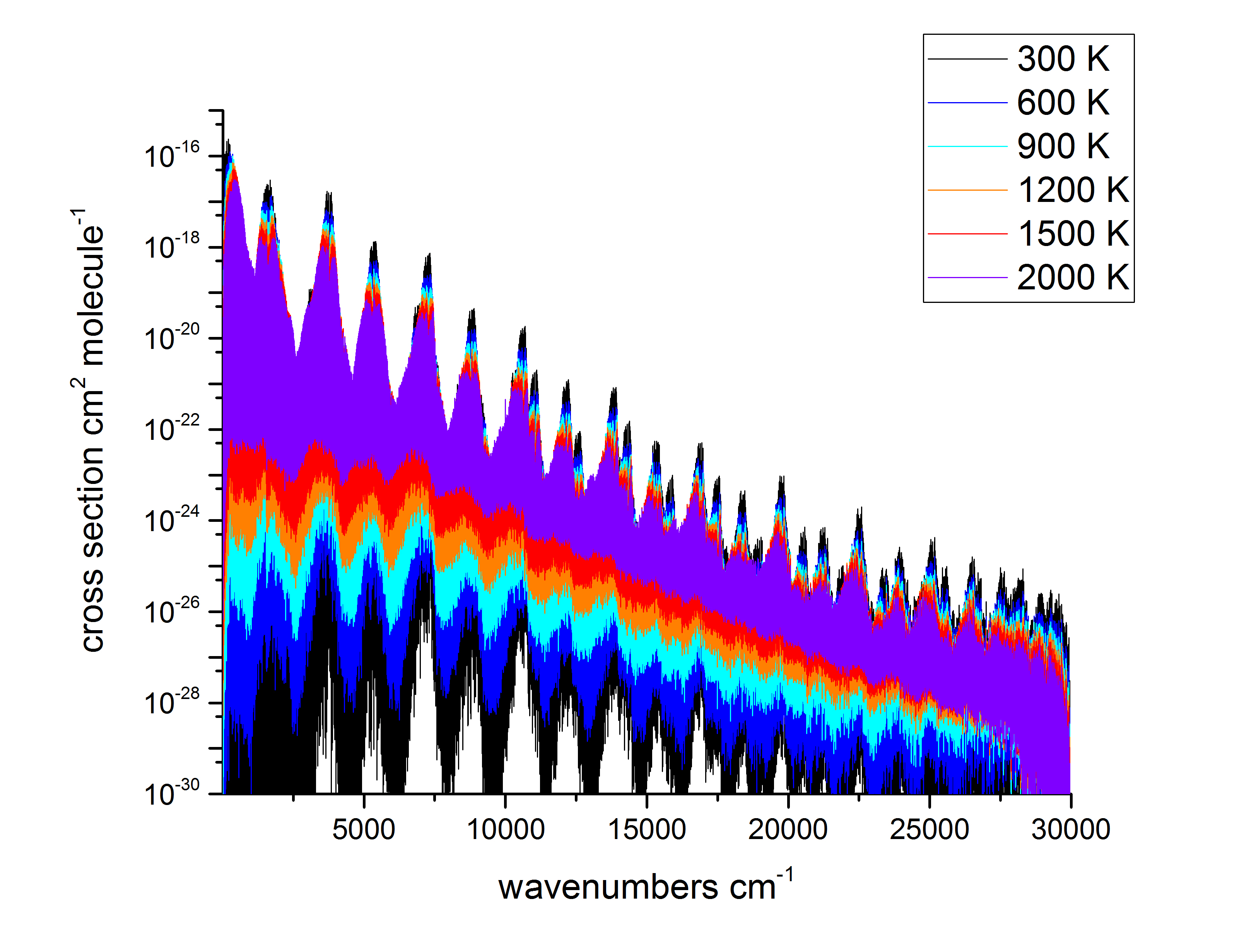}}
		\caption{\HtO\ cross sections in a mixed \Hy-He (85/15\%) environment calculated at 0.01 bar and temperatures in
			the range 300 - 2000K. A cut-off of 200 $\times (\alpha_j + \gamma_j)$ was used (see text).}
		\label{fig:Tdep}
	\end{center}
\end{figure}

\begin{figure}
	\begin{center}
		\scalebox{0.4}{\includegraphics{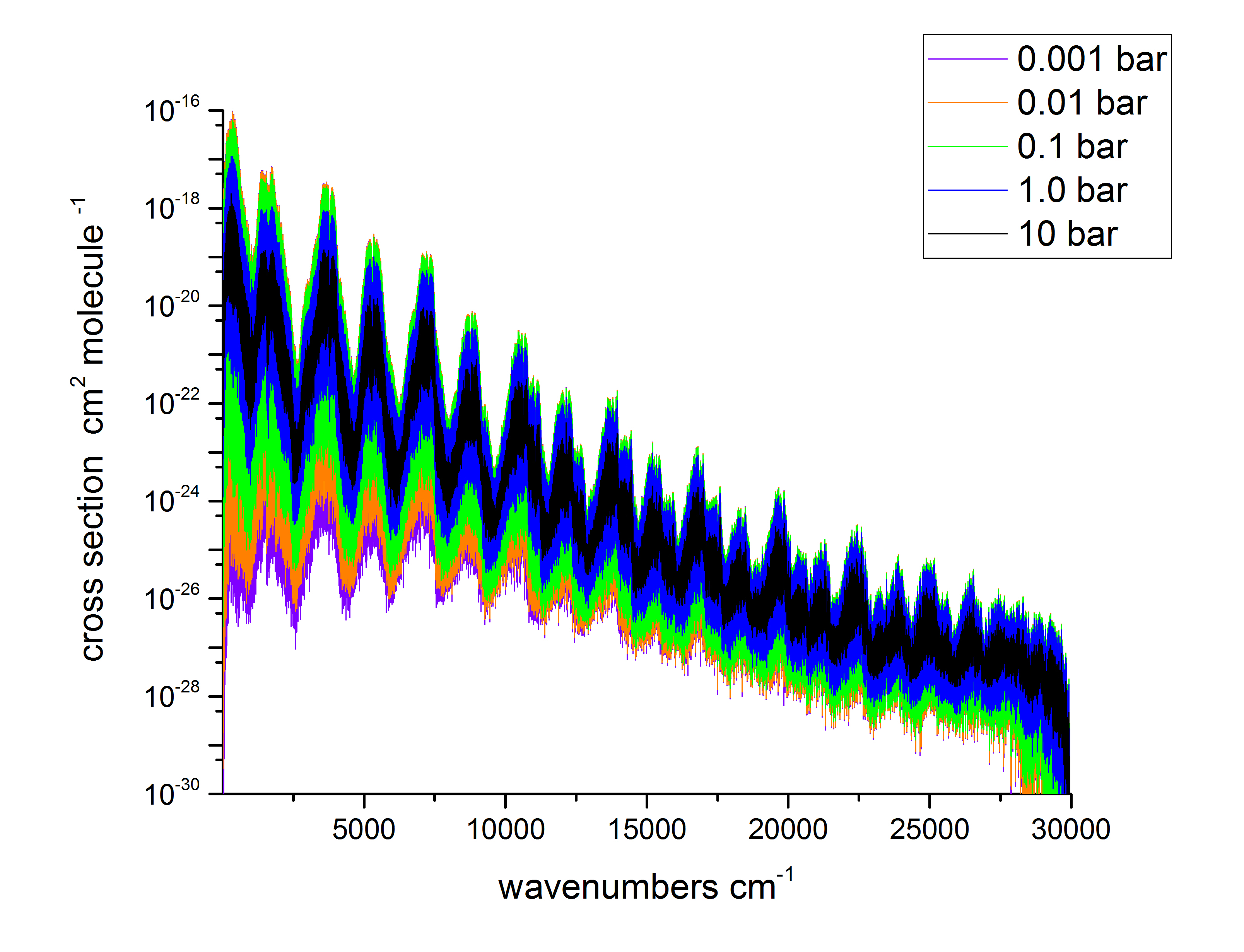}}
		\caption{\HtO\ cross sections in a mixed \Hy-He (85/15\%) environment calculated at 1000 K and pressures in
			the range 0.001 - 10 bar. A cut-off of 200 $\times (\alpha_j + \gamma_j)$ was used (see text).}
		\label{fig:Pdep}
	\end{center}
\end{figure}

Calculated cross sections for a range of temperatures at a single pressure and a
range of pressures at a single temperature are shown in Figures~\ref{fig:Tdep}
and \ref{fig:Pdep}. A change in temperature predominately influences the shape
of the cross section (Figure~\ref{fig:Tdep}) while a change in pressure results
in a redistribution of opacity (Figure~\ref{fig:Pdep}).

\begin{table}
	\caption{Temperatures and Pressures at which \HtO\ cross sections are calculated}
	\label{tab:gridTP} \footnotesize
	\begin{center}
		\begin{tabular}{cccc}
			\hline
			Temperature (K) &&& \\
			\hline
			300  &  400  & 500  & 600       \\
			700  &  800  & 900  & 1000      \\
			1200 & 1400  & 1600 & 1800      \\
			2000 &       &      &           \\
			\hline
			Pressure (bar) &&& \\
			\hline
			0.001 & 0.003 & 0.005 & 0.01    \\
		    0.02  & 0.04  & 0.08  & 0.1      \\
		    0.3   & 0.6   & 0.9   & 1.0      \\
		    1.2   & 1.5   & 2.0   & 2.5      \\
		    3.0   & 4.0   & 5.0   & 6.0      \\
		    8.0   & 10.0  &       &        \\
			\hline			
			
		\end{tabular}
	\end{center}	
\end{table}

\subsection{Interpolation of cross-sections between temperatures and pressures}

Cross sections are provided for 12 temperatures and 22 pressures between 300 - 2000 K and
0.001 and 10 bar respectively (see Table~\ref{tab:gridTP}). A cross section at intermediate
conditions may be obtained by first interpolating on the temperature
grid using either Eq. (\ref{eq:interp1}) or (\ref{eq:interp2}) below \cite{jt542}, then
interpolating on the pressure grid using Eq. (\ref{eq:interp3}) below. The interpolation
residual is expressed as a percentage of the corresponding absorption cross-section:

\begin{equation}
\delta\sigma^{\%}_{\mathrm{max}} = \mathrm{max}\left(\frac{|\sigma_{i,calc} - \sigma_{i,interp}|}{\sigma_{i,calc}}\right) \times 100
\end{equation}

\noindent
To obtain cross sections at an intermediate temperature $T$, interpolation between cross sections
computed for a higher temperature $T_2$ and a lower temperature $T_1$ may be performed linearly \cite{jt542}:

\begin{equation}
\label{eq:interp1}
\sigma_{i} = \sigma_{i}(T_1) + m(T-T_1)
\end{equation}

\noindent
where:

\begin{equation}
m = \frac{\sigma_{i}(T_2) - \sigma_{i}(T_1)}{T_2 - T_1}
\end{equation}

\noindent
Or using a more accurate exponential model \cite{jt542}:

\begin{equation}
\label{eq:interp2}
\sigma_{i} = a_i e^{\frac{b_i}{T}},
\end{equation}

\noindent
where:

\begin{equation}
b_i = \left(\frac{1}{T_2} - \frac{1}{T_1}\right)^{-1}\ln\frac{\sigma_i(T_1)}{\sigma_i(T_2)},
\end{equation}

\noindent
and:

\begin{equation}
a_i = \sigma_i(T_1)e^{\frac{b_i}{T}}
\end{equation}

\noindent
The error in the cross sections introduced by the
more accurate interpolation scheme does not exceed 1.64\% which is
less than the estimated uncertainty in the \textit{ab initio} line lists.

To obtain cross sections at an intermediate pressure $P$, interpolation between cross sections
computed for a higher pressure $P_2$ and a lower pressure $P_1$ may be performed linearly:

\begin{equation}
\label{eq:interp3}
\sigma_{i} = \sigma_{i}(P_1) + m(P-P_1)
\end{equation}

\noindent
where:
\begin{equation}
m = \frac{\sigma_{i}(P_2) - \sigma_{i}(P_1)}{P_2 - P_1}
\end{equation}

\noindent
This results in interpolation residuals below 2.6\% for the
region 6000 - 30,000 \cm\ for all pressures. However, below 6000 \cm\
for low pressures and below 1000 \cm\ for high pressures, there are
spikes in interpolation residual ($>$ 10\%) for individual wavenumber bins,
notably in the wings of strong features. The wings of strong lines
add appreciable amounts of opacity to wavenumber bins where there previously
was negligible opacity in an non-linear fashion. In reality this effects
only a small fraction of the bins ($\sim$ 0.7\%) and changes the total cross section
in the region 10 - 6000 \cm\ by less than 0.002 \%
at worst, hence this should have negligible effect on practical applications.

\section{Conclusion}

\HtO\ line widths  pressure-broadened by hydrogen and helium were
calculated using the the averaged energy difference method and $J$\p
$J$\pp-dependence technique. Rotational quantum numbers $J$ up to 50
were considered. The temperature dependence of the widths was derived
from calculations made at temperatures in the range 300 - 2000 K. The
calculated data are in reasonable agreement with available experiment.

The widths and temperature exponents are presented in new ExoMol
format as dataset exclusive \verb|.broad| files and can be used to
generate a temperature and pressure dependent Lorentzian-half width
for every line of the BT2 line list, or any line list with at least
$J$\p-$J$\pp\ assignments.

High resolution pressure dependent absorption cross sections for \HtO\
have been calculated for a mixed \Hy/He (85/15\% by number) environment and a
range of temperatures (T = 300 - 2000 K) and pressures (P = 0.001 - 10
bar) relevant to exoplanet and cool star atmospheres. The static cross
sections are available from
the ExoMol website (www.exomol.com/data/molecules/H2O/1H2-16O/BT2/).

It is out intention to make the cross sections available through the
ExoMol cross section service, a web-based interface
(www.exomol.com/data/data-types/xsec/) which currently allows astronomers to download
zero pressure cross sections for available molecules at user defined
temperatures and spectral resolution.

The form chosen for representing the pressure-broadening parameters
is based on the quantum numbers of the upper and lower levels. This
means that the parameters are transferable to other extensive water
line lists such as new,  complete and more accurate POKAZATEL line list
\cite{jtpoz} which will be released soon. Furthermore, the differences
in pressure effects between \HtO, and H$_2$$^{17}$O and  H$_2$$^{18}$O
should be very small. This means that the broadening files presented
here should also be appropriated for the newly computed
H$_2$$^{17}$O and  H$_2$$^{18}$O hot line lists \cite{jt665}

Pressure dependent absorption cross sections for other key species in
the atmospheres of exoplanets and cool stars (for example NH$_3$, CO,
CO$_2$ and CH$_4$) will also be included here in due course \cite{jtdiet}.

\section*{Acknowledgements}

This work was supported by a
grant from Energinet.dk project N. 2013-1-1027, by
UCL through the Impact Studentship Program and the European Research
Council under Advanced Investigator Project 267219 and partly supported
by CNRS in the frame of the International Associated Laboratory SAMIA and
RFBR No. 16-32-00244.

\bibliographystyle{elsarticle-num}


\end{document}